\documentclass[prl,twocolumn,aps,superscriptaddress,showpacs,floatfix]{revtex4}

\usepackage{longtable}
\usepackage{graphicx}
\usepackage{dcolumn}
\usepackage{bm}
\usepackage{color}
\begin{document}
\title{Relativistic Brownian motion on a graphene chip}

\author{Andrey Pototsky}
\affiliation{Department of Physics, Loughborough University,
Loughborough LE11 3TU, United Kingdom}
\affiliation{Department of Mathematics, University of Cape Town,
Rondebosch 7701, South Africa}
\author{Fabio Marchesoni}
\affiliation{Dipartimento di Fisica, Universit\`a di Camerino,
I-62032 Camerino, Italy}
\affiliation{Department of Physics, Loughborough University,
Loughborough LE11 3TU, United Kingdom}
\author{Feodor V. Kusmartsev}
\affiliation{Department of Physics, Loughborough University,
Loughborough LE11 3TU, United Kingdom}
\author{Peter H{\"a}nggi}
\affiliation{Institut f{\"u}r Physik
Universit{\"a}t Augsburg, D-86135 Augsburg, Germany}
\author{Sergey E. Savel'ev}
\affiliation{Department of Physics, Loughborough University,
Loughborough LE11 3TU, United Kingdom}
\begin{abstract}
Relativistic Brownian motion can be inexpensively demonstrated on a graphene chip. The interplay of stochastic and relativistic dynamics, governing the transport of charge carrier in graphene, induces noise-controlled effects such as (i) a stochastic effective mass, detectable as a suppression of the particle mobility with increasing the temperature; (ii) a transverse ratchet effect, measurable as a net current orthogonal to an ac drive on an asymmetric substrate, and (iii) a chaotic stochastic resonance. Such properties can be of practical applications in the emerging graphene technology.
\end{abstract}
 \maketitle

As astroparticle data often require a relativistic analysis, the challenge of a consistent formulation of relativistic statistical thermodynamics is so much intriguing as timely \cite{dunkel_nature}. Such a challenge encompasses the notion of relativistic Brownian motion \cite{dunkel}, as well. In particular, issues like the relativistic generalization of inherently non-local thermodynamic quantities, such as heat and work, or the very concept of an equilibrium heat bath, have not been unambiguously settled, yet  \cite{dunkel_nature,dunkel}. The state of the art in this field thus calls for experiments capable to assess and validate the existing theoretical approaches. This presently constitutes a nearly impossible task, due to the scarcity of the available relativistic data sets, which often also lack the necessary accurateness. Typically, such data either are being obtained from cosmic rays \cite{rieder} or could be, in the future, from expensive high-energy experiments. To overcome this difficulty, we propose here an alternative and affordable route to demonstrate the physics of relativistic Brownian motion under controllable laboratory conditions, namely on a suitable graphene chip.

Graphene is the only artificially crafted material known so far, which is truly two-dimensional (2D) \cite{Novoselov05}. It is essentially a monolayer of carbon atoms packed in a honeycomb lattice and isolated from the bulk. The fact that graphene has a 2D structure makes its electronic properties rather unique. Charge carriers propagating through such a lattice are known to behave as relativistic massless Dirac fermions \cite{Novoselov05,review}. On the quantum mechanical level, they are described by the 2D analog of the Dirac equation with the Fermi velocity, $v_F=10^{8}\;$cm/s, replacing the speed of light in the Dirac description. Close to the so-called Dirac points, the energy-momentum relation of the elementary excitations, often associated with quasi-particles, is linear: the velocity of the charge carriers is thus always collinear with the momentum, its modulus being constant and equal to $v_F$.

Our proposal relies on the observation that for strong applied electric fields or at high enough temperatures, quantum transport effects, otherwise dominant in graphene, are effectively suppressed and the motion of charge carriers can be satisfactorily described by classical (i.e., non-quantum) relativistic equations. Such a semiclassical description has been invoked \cite{Mikhailov09}, for instance, to successfully interpret graphene's unusually broad cyclotron resonance \cite{jiang07,deacon07}.

As detailed with this Letter, the combination of stochastic and relativistic dynamical effects governing the carrier motion in graphene results in some unexpected transport phenomena. The source of stochasticity is provided by the finite temperature, whereby the charge carriers are being scattered by phonons, lattice defects and  buckling, and the sample boundaries, which leads to an equilibrium
redistribution of their energies. We emphasize, however, that a
noisy environment in the chip can also be generated by external noise sources, controllable or not, independent of the
temperature, e.g., in the form of current or voltage fluctuations. Moreover, in graphene the Fermi velocity, which plays the role of the light speed in the relativistic equations for its charge carriers, is relatively low; one can then analyze driven relativistic Brownian motion in the presence of electromagnetic fields by ignoring otherwise hardly tractable relativistic retardation effects. 
Under these conditions, driven relativistic Brownian particles become accessible to both theory and experiment \cite{Exp-klein,wang}.

\begin{figure}[ht]
\includegraphics[width=6cm]{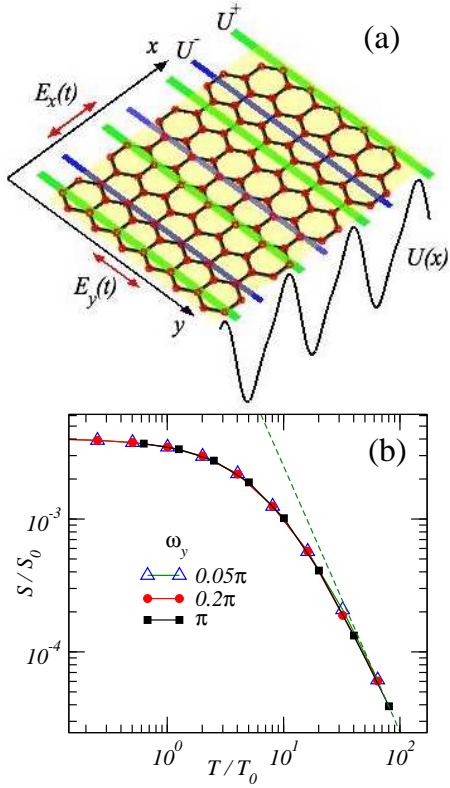}
\includegraphics[width=5.5cm]{letter_fig1b.eps}
\caption{(Color online) Panel (a): Sketch of a graphene chip (carbon atoms arranged in a honeycomb lattice) with a set of several electrodes placed parallel to the $y$-axes. The electrodes, kept at
a constant potential, $U^{+}$ (green) and $U^{-}$ (blue), create an asymmetric electric
potential, $U(x)$, shown as a solid line. External ac electric fields, $E_y(t)$ and $E_x(t)$, drive the charge carriers in the $x$ and $y$ directions, respectively.
Panel (b): Strength of the spectral peak, $S$, for a relativistic particle moving on a uniform substrate, $U(x)\equiv 0$, parallel to the ac drive $\vec{E}=(0, A\cos{(\omega_y t)})$: $S/S_0$ vs. $T/T_0$,
with $S_0=(A/\gamma\omega_y)^2$ and $T_0=v_F^2\gamma/2\omega_y$ (see text), for different $\omega_y$ (legend). Simulation parameters are: $A=20$, and $v_F=\gamma=k=1$. The dashed line represents the asymptotic decay law, $T^{-2}$, discussed in the text.
} \label{fig1}
\end{figure}
{\it Model\/} -- The scheme of a simple graphene chip is sketched in Fig.\,\ref{fig1}(a). An undoped graphene sheet sits on a periodic sequence of parallel electrodes with alternate constant potentials, respectively, $U^{+}$ and $U^{-}$, with $U^{+}>U^{-}$. 
By tuning the distance between electrode pairs within a fixed spatial period, $L$, the ensuing staggered electric potential, $U(x)$, can be modulated at will, symmetrically or asymmetrically, with amplitude $U$ in the $x$ direction, while being identically zero in the $y$ direction. An example of the directed potential $U(x)$ is drawn in Fig.\,\ref{fig1}(a). Additionally, the charge carriers are driven by a spatially homogeneous electric field, $\vec{E}(t)=(E_x(t), E_y(t))$, with sinusoidal components $E_x(t)$ and $E_y(t)$ acting along the $x$ and the $y$ axis, respectively.

The random dynamics of the charge carriers in a graphene sheet can be  described by a set of coupled Langevin equations for the components of the 2D momentum, $\vec{p}=(p_x,p_y)$, subject to the relativistic dispersion relation $\varepsilon = v_Fp_0$, where $p_0 = (p_x^2 + p_y^2)^{\frac{1}{2}}$, $\varepsilon$ is the particle energy and $v_F$ the Fermi velocity. The form of such phenomenological Langevin equations is determined by the condition that the chosen particle-reservoir coupling must lead to the same equilibrium momentum distribution as predicted by the fully microscopic (Hamiltonian) theory. At low temperatures, the equilibrium distribution function of the quasiparticle (electron) energy obeys the Fermi-Dirac statistics, $\rho(\varepsilon) = 1/(\exp{[\varepsilon/kT]}+1)$, where $T$ is the temperature and the chemical potential was set to zero to indicate that the number of the quasiparticles is undetermined. However, the operation of the graphene chips considered here typically requires carrier energies $\epsilon \sim U$, so that for sufficiently high energies, $kT \lesssim \epsilon\lesssim U$, $\rho(\varepsilon)$ is conveniently approximated by the relativistic J{\"u}ttner distribution, $\rho(\vec{p}) \sim \exp(-\varepsilon(\vec{p})/kT)$ \cite{dunkel,cuberoprl}. A viable set of Langevin equations proven to be consistent with a 2D J{\"u}ttner distribution reads \cite{dunkel}
\begin{eqnarray}\label{eq1}
\dot{p_x} &=& - \gamma v_F {p_x}/{p_0} -{dU(x)}/{dx}+ E_x(t)+\sqrt{2\gamma kT}\xi_x(t), \nonumber \\
\dot{p_y} &=& - \gamma v_F {p_y}/{p_0} + E_y(t)+\sqrt{2 \gamma kT}\xi_y(t), \nonumber \\
\vec{V} &\equiv& (\dot{x},\dot{y}) = {\partial \varepsilon}/{\partial\vec{p}}=v_F {\vec{p}}/{p_0},
\end{eqnarray}
where $\gamma$ denotes a phenomenological constant damping coefficient \cite{supplemental}. The random forces $\xi_x(t)$ and $\xi_y(t)$ are two white Gaussian noises with $\langle \xi_i(t) \rangle=0$ and $\langle \xi_i(t)\xi_j(0) \rangle=\delta_{ij}\delta(t)$, for $i,j=x,y$, which ensures proper thermalization at temperature $T$. We recall here that the modulus of $\vec{V}$ is constant and equal to $v_F$ \cite{Mikhailov09}. These equations describe a relativistic Brownian dynamics, thus suggesting new settings for the experimental and theoretical investigation of relativistic thermodynamics and relativistic Brownian motion. Below we report results from extensive numerical simulations we performed by integrating Eqs. (\ref{eq1}). 

%
As anticipated above, the validity of Eqs. (\ref{eq1}) for electrons in graphene is restricted to the quasi-classical limit, where quantum mechanical effects can be safely neglected. Thus, the testing ground for relativistic phenomena on a graphene chip is subject to the  following physical restrictions  \cite{supplemental}:\\
(i) the de Broglie wavelength across the barrier, $\lambda_x=\hbar/p_x$, must be much smaller than the period of substrate potential, $\lambda_x \ll L$, so as to neglect either miniband or discrete energy levels;\\
(ii) the de Broglie wavelength parallel to the barrier, $\lambda_y=\hbar/p_y$, must be much smaller than $L$, so as to suppress the probability of chiral tunneling (Klein paradox \cite{Exp-klein,katsnelson_06}). The condition $\lambda_y \ll L$ allows neglecting, in particular, the electron-hole tunneling;\\
(iii) temperature (or noise strength) must be high enough to further ensure that chiral tunneling at the top of the $U(x)$ barriers is negligible with respect to the competing noise-activated particle hopping, i.e.,
$$\min\{{\hbar v_F}/{L}; {\hbar \omega_{y}^{2}U^2}/{v_F A^2L}\}\ll kT \lesssim U.$$

Condition (iii) on $T$ is introduced \cite{supplemental} because noiseless carriers moving along the $x$-axis with $\lambda_y\ll L$ would still be subject to Klein's mechanism (barrier transparency) \cite{Efetov}. Note, however, that in 2D geometries with smooth potentials $U(x)$, the fraction of electrons undergoing Klein's mechanism is negligibly small, not only at high temperatures [as guaranteed by condition (iii)], but also in the presence of sufficiently strong transverse drives. Indeed, drives oriented along the $y$ axis tend to suppress chiral tunneling even for particles with $p_y=0$ \cite{efetov-optics}, thus corroborating our quasiclassical model.

\begin{figure}[tp]
\includegraphics[width=6cm]{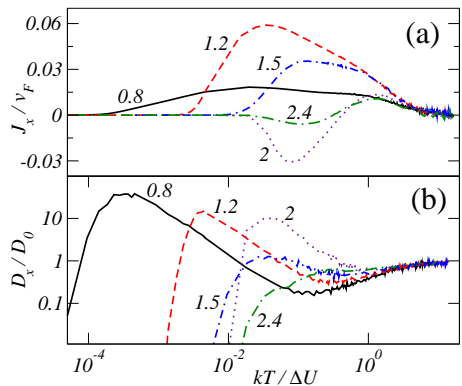}
\caption{(Color online) Relativistic ratcheting in an asymmetric, periodic potential $U(x)=\sin{(2\pi x)}+0.25\sin{(4\pi x)}$, driven by $\vec{E}=(0, A\cos{(\omega_y t)})$: (a)
Net current $\bar{J}_x/v_F$ versus $T$ for different $\omega_y$
(reported next to each curve); (b) Rescaled diffusion coefficient, $D_x/D_0$, versus $T$ for the same simulation parameters as in (a).
Here, $D_0=kT/\gamma$ and $\Delta U\simeq 2.20$ (height of the substrate barriers).
Other parameters are $A=20$ and $v_F=\gamma=k=1$.
} \label{fig2}
\end{figure}
{\it Stochastic effective mass} --  The moment components $p_x$ and $p_y$ in Eqs. (\ref{eq1}) are coupled via the dispersion relation, $\varepsilon = v_Fp_0$, even in the
{\it absence} of a substrate potential, 
$U(x)\equiv 0$. This gives rise to a peculiar phenomenon, which
characterizes the relativistic Brownian motion with respect to its non-relativistic counterpart [Fig. \ref{fig1}(b)]. By numerically solving Eqs. (\ref{eq1}) for electrons driven, say, along the $y$ axis by $\vec{E}=(0, A\cos{(\omega_y t)})$, we determined the strength, $S$, of the delta-like peak of the $y$ power spectral density at the driving frequency, $\omega_y$. Contrary to the non-relativistic limit, where it is $T$ independent, at high temperatures $S$ decreases according to the approximate power law $T^{-2}$. This result can be quantitatively explained by noticing that according to the high-$T$ J\"uttner distribution, the particle energy distribution is $\rho(\epsilon) \sim \epsilon \exp(-\epsilon/kT)$; hence, $\langle \epsilon \rangle = v_F \langle p_0 \rangle = 2kT$. Upon replacing $p_0$ by $2kT/v_F$, Eq. (\ref{eq1}) is reduced to a set of decoupled Langevin equations describing a non-relativistic particle of effective mass $m_{\rm eff}=2kT/v_F^2$. Accordingly, $S/S_0=1/[1+(\omega_y/\omega_0)^2]$ with $S_0=(A/\gamma \omega_y)^2$ and $\omega_0 =\gamma/m_{\rm eff}=v_F^2\gamma/2kT$ [dashed line in Fig. \ref{fig1}(b)]. Stochastic mass renormalization provides a simple validation check for our phenomenological model in Eqs. (\ref{eq1}).

{\it Relativistic ratcheting\/} -- We consider next charge carrier
transport due to the rectification of non-equilibrium perturbations on a substrate with asymmetric potential $U(x)$  (ratchet or Brownian motor effect \cite{hanggi_march_09}).
The working principle of a Brownian motor is that, under certain
conditions, asymmetric devices are capable of rectifying random
(i.e. noisy) and/or deterministic (periodic) modulations. In the non-relativistic regime, ratcheting occurs only in the $x$ direction, as a mere effect of the $x$ component of the periodic drive, $E_x(t)$. The rectification current weakens if $\vec{E}(t)$ is rotated at an angle with the $x$-axis, until it drops to zero for ac drives parallel to the substrate valleys. Indeed, the component $E_y(t)$ of the ac-field does keep the system out of equilibrium, but cannot be rectified, since the substrate potential is uniform in the $y$ direction. Stated otherwise, the $x$ and $y$ dynamics are decoupled.
In the relativistic Langevin equations (\ref{eq1}), instead, the orthogonal ac-drive components, $E_x(t)$  and
$E_y(t)$, are nonlinearly coupled through the dispersion relation, so that both can be rectified by the asymmetric potential $U(x)$.
Most remarkably, apart from a special parameter range dominated by
relativistic chaotic effects (see below), transverse rectification induced by $E_y(t)$ in the $x$ direction only occurs at finite temperatures, thus implying {\it a bona fide} noise-sustained Brownian ratchet \cite{hanggi_march_09}. Noise is required to force particle fluctuations in the $x$-direction, around the asymmetric minima of $U(x)$; as $p_y$ is driven by $E_y(t)$ toward values of the order of $\sqrt{T}$, or smaller, correspondingly $v_x$ jumps to a maximum, $v_x \approx v_F p_x/|p_x|=\pm v_F$, so that the particle is kicked in the $x$ direction, either to the right or to the left; this is how spatial asymmetry comes into play \cite{movie}. Naturally, such a rectification mechanism becomes ineffective when the particle sits at a potential minimum with $p_x=0$, which only occurs in the absence of noise, i.e., for $T\equiv0$.

\begin{figure*}[ht]
\includegraphics[width=17.0cm]{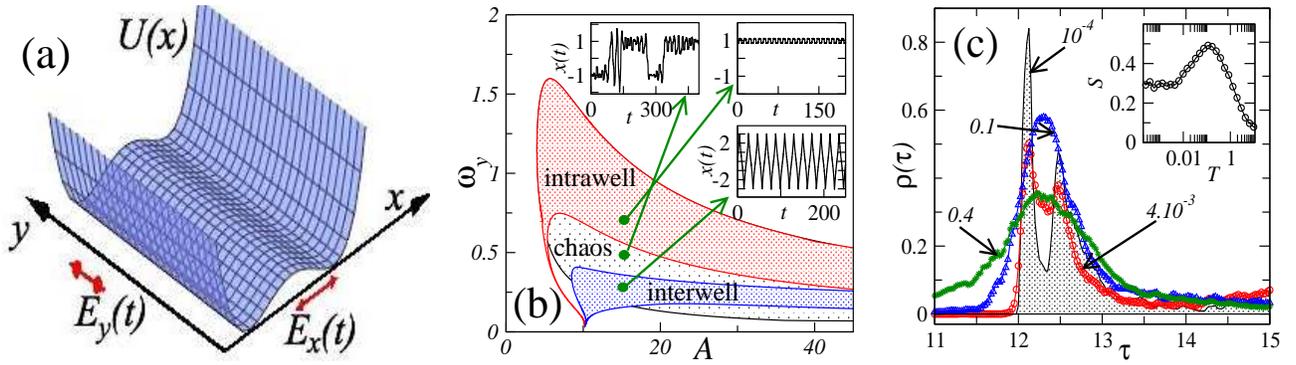}
\caption{(Color online) Relativistic stochastic resonance. (a) The confining potential, $U(x)=x^4/4 -x^2/2$, bistable in the $x$-direction and uniform in the transverse direction $y$;
(b) Corresponding phase diagram for $\vec{E}=(0, A\cos{(\omega_y t)})$. Insets: trajectory samples for points in the plane $(A,\omega_y)$ (denoted by arrows) belonging to different dynamical phases. (c) Residence time distributions, $\rho(\tau)$, for different $T$ (as indicated) and constant drive, $A=15$ and $\omega_y=0.5$. Relevant simulation parameters: integration step, $1.2 \times 10^{-3}$; run time length, $5 \times 10^5$; distribution time bin, $2.5 \times 10^{-2}$. Inset: the distribution peak strength, $S=\int_{11}^{14} \rho(\tau) d\tau$, plotted versus $T$ (SR signature). Other simulation parameters are: $v_F=\gamma=k=1$.
} \label{fig3}
\end{figure*}
To demonstrate the relativistic transverse ratchet effect,
it suffices to arrange the electrodes on the graphene chip so as to generate an asymmetric periodic potential, $U(x)$, directed along $x$ and ac drive the charge carriers along $y$ with $\vec{E}(t)=(0, A\cos \omega_y t)$. By numerically integrating the corresponding Eqs. (\ref{eq1}), we observed that a relativistic particle tends, indeed,
to drift in the $x$ direction with net current $\bar{J}_x
= \langle \dot{x}\rangle$ and diffusion coefficient $D_x =
\lim_{t\rightarrow \infty}[\langle x(t)^2\rangle -\langle
x(t)\rangle^2]/2t$. The dependence of $\bar{J}_x$ and $D_x$ on the
temperature and the driving frequency are displayed in Fig.\, \ref{fig2}. In panel (a) it is apparent that at constant $\omega_y$ the absolute value of the current, $\mid \bar{J}_x \mid$, hits a maximum for an optimal $T$ and tends to zero in the limits $T\to 0$ and $T\to \infty$, thus underscoring the key role played by noise. This result supports our conclusion that we deal with a new category of noise-sustained (no  current at zero noise) and relativistic (no ratchet phenomenon in the non-relativistic limit) Brownian motors. The direction of the net rectification current depends on both the driving frequency and the temperature; its sign can be (multiply) reversed by changing either $T$, panel (a), or $\omega_y$ (not shown).
Most remarkably, in panel (b) one can locate a finite temperature interval, where, for a given $\omega_y$, the scaled diffusion coefficient $D_x/D_0$ decreases with increasing $T$; indeed, $D_x$ approaches the expected asymptotic value $D_0=kT/\gamma$ only after going through a minimum. This $T$-interval approximately coincides with the $T$-interval in panel (a), where the absolute value of the
current and, therefore, the mobility of the charge carriers is the
largest. Analogously, we checked that the absolute maximum of $D_x$ as a function of $\omega_y$ corresponds to a local minimum of $\bar{J}_x$ versus $\omega_y$. This observation suggests that for appropriate system parameters, elementary charge excitations in graphene can be regarded as Brownian particles with large mobility and low diffusivity.

{\it Chaos and stochastic resonance} -- To investigate the role of anomalous signal amplification we consider relativistic Brownian motion driven along $y$ by the time-periodic electric field $\vec{E}(t)=(0,A\cos(\omega_y t))$, and confined along $x$ by the double-well potential $U(x)=x^4/4 - x^2/2$, sketched in Fig.\,\ref{fig3}(a). In the deterministic, or noiseless regime, $T\equiv 0$, the motion of a {\it classical} particle in the
$x$-direction would be frozen, i.e. the particle would sit in one of the two $U(x)$ minima, regardless of the drive applied in the $y$ direction. In clear contrast, the response of the relativistic particle in the $x$-direction is strongly affected by the nonlinear coupling of the $x$- and $y$-dynamics, controlled, respectively, by $U(x)$ and  $E_y(t)$. In fact, by appropriately tuning the drive parameters $A$ and $\omega_y$, the relativistic particle can execute either small amplitude intra-well oscillations within one confining well, or large-amplitude inter-well oscillations between the two confining wells, or even exhibit chaotic-like switching between the two wells. The phase diagram of these three different dynamical regimes is depicted in Fig.\,\ref{fig3}(b) with samples of the corresponding  $x$-trajectories reported in the insets.

Our numerical simulations showed that in the chaotic regime a relativistic particle performs small amplitude oscillations around one $U(x)$ minimum for an unpredictable time interval (residence time) before switching over into the other minimum, where it resumes its small amplitude oscillations until the next switching
event. These switchings are due to the intrinsic instability of the particle dynamics and {\it not} to thermal fluctuations, as in our simulations the temperature was initially set to zero.

On raising the temperature, the inter-well dynamics induced by the transverse field, $E_y(t)$, becomes increasingly noise dominated. In Fig.\,\ref{fig3}(c) we display the normalized distribution density of the residence times in either $U(x)$ well, $\tau$,
at different $T$. The peak split structure, characteristic of the noiseless chaotic dynamics, merges into a single broad peak centered at around $\tau=2\pi/\omega_y$. The corresponding peak strength, $S$, defined here as the area encircled by the peak, attains a maximum for $kT$ smaller than, but close to the barrier height of $U(x)$, $\Delta U=0.25$. Such a behavior can be regarded as an instance of stochastic resonance, a phenomenon well established for bistable non-relativistic Brownian motion \cite{March98}, with two additional peculiarities: Here the periodic drive acts perpendicularly to the hopping direction and the resonance is related with chaotic rather than periodic switches between potential wells.

We conclude by remarking that, irrespective of the phenomenological details of Eqs. (\ref{eq1}), the spatial coordinates of a relativistic particle in 2D are nonlinearly coupled via a non-separable energy-momentum dispersion relation. In this regard our approach may apply not only to graphene but also to other materials, like certain types of semiconductors. Moreover, the results discussed in this work, besides suggesting an inexpensive technique to assess the current stochastic formalisms for relativistic particles, might well find application in ultrafast electronics. Indeed, pursuing the technological implementation of the effects reported above (and  yet others as summarized in Ref. \cite{supplemental}) is expected to become viable, should one eventually succeed in replacing GaAs- with graphene-based electronics \cite{Geim07}.

\acknowledgements {Work partially supported by the European Commission (grant No. $256959$, NanoPower), the Alexander von Humboldt Foundation, the EPSRC (EP/D072581/1), the German excellence cluster ``Nanosystems Initiative Munich'' (NIM) and the Augsburg center for Innovative Technology (ACIT) of the University of Augsburg.}

\end{document}


\title{Supplemental material for ``Relativistic Brownian motion on a graphene chip''}

\author{Andrey Pototsky}
\affiliation{Department of Physics, Loughborough University,
Loughborough LE11 3TU, United Kingdom}
\affiliation{Department of Mathematics, University of Cape Town,
Rondebosch 7701, South Africa}
\author{Fabio Marchesoni}
\affiliation{Dipartimento di Fisica, Universit\`a di Camerino,
I-62032 Camerino, Italy}
\affiliation{Department of Physics, Loughborough University,
Loughborough LE11 3TU, United Kingdom}
\author{Feodor V. Kusmartsev}
\affiliation{Department of Physics, Loughborough University,
Loughborough LE11 3TU, United Kingdom}
\author{Peter H{\"a}nggi}
\affiliation{Institut f{\"u}r Physik
Universit{\"a}t Augsburg, D-86135 Augsburg, Germany}
\author{Sergey E. Savel'ev}
\affiliation{Department of Physics, Loughborough University,
Loughborough LE11 3TU, United Kingdom}
 \maketitle

\section {Quasi-classical dynamics of charge carriers in graphene}

We discuss here our arguments in support of the semiclassical description, Eqs. (1), main text, of the carrier dynamics in a realistic graphene sheet coupled to a heat-bath at given temperature $T$.

In the low-energy limit and in the absence of randomness (fluctuations, $T\equiv 0$, and disorder, alike), the charge carriers in graphene, electrons and holes, obey the Dirac equations,
\begin{eqnarray}
\label{newd}
({\cal P}_x-i{\cal P}_y)\Psi_B+\frac{{\cal U}(x,y,t)}{v_F}\Psi_A=\frac{i\hbar}{v_F}\frac{\partial \Psi_A}{\partial t}\\ \nonumber
({\cal P}_x+i{\cal P}_y)\Psi_A+\frac{{\cal U}(x,y,t)}{v_F}\Psi_B=\frac{i\hbar}{v_F}\frac{\partial \Psi_B}{\partial t},
\end{eqnarray}
where the two spinor components of the carrier wave function,
$(\Psi_A, \Psi_B)$, refer to the two triangular graphene sublattices and the momentum operators are defined as
$({\cal P}_x,{\cal P}_y)=(-i\hbar\partial/\partial x,-i\hbar\partial/\partial y)$.

Applying the operators $({\cal P}_x-i{\cal P}_y)$ and $({\cal P}_x+i{\cal P}_y)$, respectively, to the first and the second equation  leads to the coupled second-order partial differential equations,
\begin{eqnarray}
\left[-\hbar^2\left(\frac{\partial^2}{\partial x^2}+\frac{\partial^2}{\partial y^2}\right)+\frac{\hbar^2}{v_{F}^{2}}
\frac{\partial^2}{\partial t ^2}+2\frac{i\hbar}{v_{F}^{2}}{\cal U}\frac{\partial}{\partial t}-\frac{{\cal U}^2}{v_{F}^{2}}\right]\Psi_B=
-\frac{i\hbar}{v_{F}^{2}}\Psi_B\frac{\partial{\cal U}}{\partial t}+\frac{i\hbar}{v_{F}}\Psi_A\left(\frac{\partial{\cal U}}{\partial x}+i
\frac{\partial{\cal U}}{\partial y}\right),\\ \nonumber
\left[-\hbar^2\left(\frac{\partial^2}{\partial x^2}+\frac{\partial^2}{\partial y^2}\right)+\frac{\hbar^2}{v_{F}^{2}}
\frac{\partial^2}{\partial t ^2}+2\frac{i\hbar}{v_{F}^{2}}{\cal U}\frac{\partial}{\partial t}-\frac{{\cal U}^2}{v_{F}^{2}}\right]\Psi_A=
-\frac{i\hbar}{v_{F}^{2}}\Psi_A\frac{\partial{\cal U}}{\partial t}+\frac{i\hbar}{v_{F}}\Psi_B\left(\frac{\partial{\cal U}}{\partial x}-i
\frac{\partial{\cal U}}{\partial y}\right).
\end{eqnarray}
Equivalently, the effective actions $(S_A, S_B)$, defined by rewriting $(\Psi_A, \Psi_B)$ as $(\exp(iS_A/\hbar),\exp(iS_B/\hbar))$, satisfy the coupled equations,
\begin{eqnarray}
\left(\frac{\partial S_A}{\partial x}\right)^2+\left(\frac{\partial S_A}{\partial y}\right)^2 -\frac{1}{v_{F}^{2}}\left(\frac{\partial S_A}{\partial t} - {\cal U}\right)^2 -i\hbar\frac{\partial^2 S_A}{\partial x^2} - i\hbar\frac{\partial^2 S_A}{\partial y^2} +\frac{i\hbar}{v_{F}^{2}}\frac{\partial^2 S_A}{\partial t^2}=-\frac{i\hbar}{v_{F}^{2}} \frac{\partial U}{\partial t}+\frac{\hbar}{v_F}e^{\frac{i}{\hbar}(S_A-S_B)}\left(i\frac{\partial {\cal U}}{\partial x}-\frac{\partial {\cal U}}{\partial y}\right), \nonumber \\
\left(\frac{\partial S_B}{\partial x}\right)^2+\left(\frac{\partial S_B}{\partial y}\right)^2 -\frac{1}{v_{F}^{2}}\left(\frac{\partial S_B}{\partial t} - {\cal U}\right)^2 -i\hbar\frac{\partial^2 S_B}{\partial x^2} - i\hbar\frac{\partial^2 S_B}{\partial y^2} +\frac{i\hbar}{v_{F}^{2}}\frac{\partial^2 S_B}{\partial t^2}=-\frac{i\hbar}{v_{F}^{2}} \frac{\partial U}{\partial t}+\frac{\hbar}{v_F} e^{\frac{i}{\hbar}(S_B-S_A)}\left(i\frac{\partial {\cal U}}{\partial x}+\frac{\partial {\cal U}}{\partial y}\right).\nonumber
\label{jacobi}
\end{eqnarray}
In the limit $\hbar\rightarrow 0$, $S_A$ and $S_B$ tend to the unique classical action, $S$, solution of the classical Hamilton-Jacobi equation,
\begin{equation}
\left(\frac{\partial S}{\partial x}\right)^2+\left(\frac{\partial S}{\partial y}\right)^2 -\frac{1}{v_{F}^{2}}\left(\frac{\partial S}{\partial t} - {\cal U}\right)^2 = 0.
\label{h-jacobi}
\end{equation}
On making use of the well-known relation, ${\cal H}=\partial S/\partial t$, one eventually derives from Eq. (\ref{h-jacobi}) the classical Hamiltonian for a relativistic massless particle,
\begin{equation}
{\cal H}=\pm v_F\sqrt{p_{x}^2+p_{y}^{2}}+{\cal U},
\end{equation}
with ${\cal U}(x,y,t)= U(x)-E_y(t)y-E_x(t)x$ and $\pm$ referring respectively to electrons and holes, whose densities are related by electroneutrality.

Note that the lowest order corrections to the classical limit, Eq. (\ref{h-jacobi}), are ${\cal O} (\hbar)$, at variance with the standard WKB formalism for the Schr\"odinger equation. Such  corrections, for $|S_A-S_B| \sim \hbar$, are due to the electron-hole transitions, so that a systematic quasiclassical treatment would require a two particle species description.

Thus far, the interactions of a single charge carrier with other carriers, phonons and impurities have been neglected. If the relevant collision times are short compared with the operation time constants of the graphene chip (periods of the ac-drives, carriers drift times across the substrate potential cells, etc.), then such interactions can be accounted for by adding to the potential function, ${\cal U}$, an appropriate fluctuating term \cite{tsvelick}, $U_{\xi}$,  namely, ${\cal U}(x,y,t) = U(x)-E_y(t)y-E_x(t)x + U_{\xi}$, which corresponds to an isotropic white noise force in the plane $(x,y)$.

The same interactions are also responsible for the damping term in the Langevin equations (Eq. (1), main text). The appropriate form of the damping and the fluctuating terms are determined by the fluctuation-dissipation theorem. In Eqs. (1), main text, we first introduced a damping term of the viscous type, $-\gamma {\vec V}$ \cite{mallick}, and then determined the random forces accordingly \cite{dunkel}. Such a choice is in line with the single $\tau$-approximation of the Boltzmann collision integral, which, for systems not too far from equilibrium, naturally yields a viscous term with $\gamma=1/\tau$, see, e.g., in Ref. \cite{kiril}.

The damping coefficient $\gamma$ contains contributions from all possible collision mechanisms a charge carrier may undergo when moving through a graphene chip. To determine the relative importance of such mechanisms would require a detailed quantum kinetic theory (see, e.g. Ref.  \cite{kinetic}), which rests beyond the purposes of our semi-classical approach, and, more importantly, direct access to experimental parameters, such as spatial disorder (impurities), internal lattice stress in the presence of a substrate (phonon distribution), local temperature and magnetic field fluctuations, which are hardly controllable. Recent experimental evidence \cite{filleter,seol} seems to indicate that the main source of carrier viscosity in graphene is attributable to electron-phonon interactions. Anyway, at the present stage, $\gamma$ ought to be regarded as a phenomenological parameter of the semiclassical Eq. (1), main text, to be experimentally determined.

The semiclassical approach outlined above provides a meaningful phenomenological description of charge carrier transport in a graphene chip as long as its characteristic length scales (e.g., the chip size and the electrode spacings) are much larger than the de Broglie wavelengths, $\lambda_x=\hbar/p_x$ and $\lambda_y=\hbar/p_y$, with $(p_x, p_y)=(\partial S/\partial x, \partial S/\partial y)$, see conditions (i) and (ii) in the main text. Condition (i) states that for $\lambda_x$ much shorter than the characteristic length scale of the substrate potential along the $x$-axis, $L\sim |U/(dU/dx)|$ for smooth functions $U(x)$, the discreteness of the carrier energy spectrum becomes negligible, as to be expected in view of the Bohr-Sommerfeld quantization rule.

Condition (ii), $\hbar/p_y\ll L$, follows from requiring that the probability of chiral tunneling is distinctly smaller than unity, i.e., that the particle must be reflected many times by the potential barrier before it may tunnel across. Chiral tunneling is exponentially suppressed for non-zero transverse momenta, $p_y\ne 0$, and smooth time-independent potentials, $U(x)$. In this case, the classical action has the form $S=p_y y-Et+S_0(x)$, where $E$ is the conserved energy and $S_0(x)$ can be readily
derived from the the Hamilton-Jacobi Eq. (\ref{h-jacobi}), that is, $S_0=\pm\int\sqrt{(E-U)^2/v_{F}^{2}-p_{y}^2}dx$.
Chiral tunneling occurs through the classically forbidden region,
$|E-U|<v_F|p_y|$; therefore, the smoother the potential, the larger the size of the forbidden region, $2v_F|p_y|/|dU/dx|$. As a consequence,
chiral tunneling is characterized by an exponentially small probability, $w=\exp(-\zeta_0 v_{F}p_{y}^{2}/\hbar |dU/dx|)$ \cite{Efetov},
with $\zeta_0$ denoting a constant of the order of unity, which may be appreciable only for a small fraction of electrons with $p_y\lesssim \sqrt{\hbar |dU/dx|/v_{F}}$. By introducing the estimate $p_y\sim U/v_F$ for the transverse momentum, we derived condition (ii). It is important to stress that a time dependent electric field acting along the barrier, $E_y(t)$, tends to exponentially suppress also the chiral tunneling of electrons moving perpendicularly to the barrier, $p_y=0$, that is, $w=\exp[-\pi e^2v_FA^2/(4\hbar|dU/dx|\omega_{y}^2)]$  \cite{efetov-optics}. This remark further corroborates our quasiclassical approach in the presence of strong enough drives.

As an additional consistency requirement for our semiclassical description of graphene charge carrier transport over the substrate potential, we must further ensure that chiral barrier tunneling is negligible with respect to the thermal barrier hopping, that is, $w \ll \exp(-U/kT)$. This yields the lower bound  for temperature in condition (iii).

In conclusion, our quasiclassical Langevin equations, Eqs. (1), main text, represent a valid starting point to model carrier transport in graphene at finite temperature (see e.g., \cite{kittel}), with one proviso: Such a phenomenological description only applies as long as (i) chiral tunneling (which implies the conservation of the associated chiral quantum number), (ii) interband transitions, and (iii) intervalley scattering can safely be ignored.
As mentioned above, such effects could be accounted for, at least in principle, by developing a more refined phenomenological (semiclassical) model, which involves two (or more) species \cite{ourprls} of interacting particles, electrons and holes, and  thus is capable to describe an even richer relativistic stochastic dynamics.

\section{Relativistic vs. non-relativistic Brownian motion}
The new effects observed so far are summarized in Table\,\ref{tab}, where for reader's convenience we compare case by case the different behaviors of relativistic (massless) and non-relativistic (massive, or overdamped) Brownian particles confined to a two dimensional geometry. One has to bear in mind that the
orthogonal coordinates of the relativistic particle described by the Eqs. (1), main text, are dynamically coupled even in the absence of a nonlinear substrate or external gradients. Such a coupling vanishes in the non-relativistic limit. As a consequence, first, the response of graphene charge carriers to a time-periodic drive is temperature dependent, see Fig.1b, whereas the response of a classical free particle only depends on the drive. Second, on applying a periodic symmetric spatial modulation in a given direction, a relativistic massless particle can be rectified in that direction,
when driven by two center-symmetric unbiased time periodic fields (harmonic mixing), acting in orthogonal directions. Note that our data on harmonic mixing are not reported in the manuscript. Rectification occurs only for a certain combination of the relevant driving frequencies \cite{archiv}.
In the case of a classical particle, harmonic mixing would require that both harmonic signals act in the direction of the spatial modulation. Third, in the presence of a spatially asymmetric ratchet-like potential, $U(x)$ (Fig. 2), the relativistic
particle can be rectified along $x$ by periodically forcing it along $y$, contrary to classical ratchets, where rectification of transverse signals was never observed.
Finally, if the motion of a noiseless relativistic particle is confined to a double-well potential (Fig. 3) along $x$, in the presence of a transverse harmonic drive, $E_y(t)$, its coordinate $x(t)$ can be either constant, or periodic, or chaotic, depending on the $E_y(t)$ parameters.
Adding noise causes an unusual manifestation of stochastic resonance with no counterpart in the non-relativistic regime: at finite temperature, particle switchings between $U(x)$ minima can be optimally synchronized even by means of a transverse periodic modulation.

This work was partly supported by the HPC-Europa2 Transnational Access Programme, application N 278; European Commission (NanoPower project); the Alexander von Humboldt Foundation, EPSRC (EP/D072581/1) and the German excellence
cluster''Nanosystems Initiative Munich'' (NIM) via its seed funding program.
\begin{table}[ht]
\caption{\label{tab} Driven stochastic dynamics of relativistic and
non-relativistic particles: A comparison}
\begin{ruledtabular}
\begin{tabular}{lll}
\hline
&&\\
& Relativistic massless particles & Classical massive or \\
& & overdamped particles \\
&&\\
\hline
&&\\
Experimental data sources   & Cosmic rays, graphene chip                            & nano-particles in liquids\\
                            &                                                       & and nano-channels,\\
                            &                                                       & vortices in superconductors, \\
                            &                                                       & bio-molecules in bio-systems etc.\\
&&\\
\hline
&&\\
Possible applications       & THz graphene-based electronics                        & nanotechnology,\\
                            &                                                       & noise-driven nano-robots,\\
                            &                                                       & drug delivery etc.\\

&&\\
\hline
&&\\
Current state of the art    & conflicting formulations                                 & well established working tool\\
                            & limited applicability                                                      & \\

&&\\
\hline
&&\\
Coupling between orthogonal coordinates   & Yes                            & No \\
&&\\
\hline
&&\\
Rectification in a spatially symmetric    & Occurs if the signals are      & Occurs only if the two signals are\\
periodic potential $U(x)$, induced        & applied in the same as well as & applied in the same direction \\
by mixing of two harmonic signals         & in the orthogonal directions   &  \\
&&\\
\hline
&&\\
Rectification in a spatially asymmetric   & Occurs if the signal is applied along       & Occurs only if the signal \\
periodic potential $U(x)$ induced by      & the $x$, as well as along the $y$ direction & is applied along the $x$ direction   \\
a single center-symmetric unbiased signal &                                             &   \\
&&\\
\hline
&&\\
Motion in a double-well potential         &  Zero temperature lateral response          & No zero temperature lateral response \\
with orthogonal harmonic drive;           &  can be constant, periodic or chaotic     & \\
chaos and resonance                       &   in time.                                         &\\
                                          &  Finite temperature response in the               & Finite temperature response: random\\
                                          &  chaotic regime: non-exponential                  & switchings with exponential distribution\\
                                          &  double-peak distribution of the                  & of the residence times\\
                                          &  residence times; stochastic resonance                     &\\
\hline
\end{tabular}
\end{ruledtabular}
\end{table}